\documentclass[11pt,epsf]{article}
\usepackage{graphicx}

\begin{document}
\sloppy
\author{Vladim\'{\i}r Majern\'{\i}k,\\
Institute of Mathematics, Slovak Academy of Sciences, \v
Stef\'anikova 49,
\\SK-814 73 Bratislava, Slovak Republic}
\title{ Energy Conservation at the
Gravitational Collapse}
\maketitle
\begin{abstract}
We apply the principle of energy conservation to the motion of the
test particle in gravitational field by requiring that its energy,
gained by gravitation, has to be balanced by decrease of its rest
mass. Due to the change of mass in gravitational field Newton's
force law between gravitating bodies is modified, too. With this
modified force law we build up the the classical field theory of
gravitation in which all relevant field quantities are in the
definition domain $r\in [0,\infty)$ finite and positive. We show
that under such circumstances, the energy release at any
gravitational collapse is finite. On the other side, the energy
conservation leads to an equation which relates the mass change of
the test particle due to  gravitation and the metric of the
corresponding gravitational field. The mass change in Newton's
gravitational field lead to a remarkable simple metric which shifts,
in contrast to the Schwarzschild metric, the horizon of events to
the gravity center of the gravitational collapse.\\


\end{abstract}
\section{Introduction}
It is generally accepted that if a star-like object with a
sufficient high mass is collapsing over a certain limit radius, it
becomes unstable and due to stronger and stronger gravitation forces
it shrinks to a mathematical point. In Newtonian physics the final
state of such gravitational collapse - the naked singularity - is a
state of infinite mass density not separated from the outside word
by a horizon of events and its formation is accompanied by radiating
an infinite amount of heat. The general relativity offers the
possibility of stable end state of collapsing mass body called black
hole representing a mass object from which elm radiation can not
escape. Beside the general feeling of physicists who have a
psychological resistance against the existence of any singular state
of matter in both cases one also faces difficulties with the
physical interpretation of the final product of collapsing star. The
classical naked singularity represents a highly unphysical state of
matter and the existence of black hole in its general form is
conditioned with the cosmic censorship hypothesis which appears not
generally prove so far. Therefore, the endeavor to find alternatives
to the classical naked singularity as well as to the black holes
seems still to be justified. In what follows, we will investigate
the possibility of creation of the naked singularity and the
classical black hole by strict application principle of energy
conservation to the gravitational interaction.

As is well-known, the {\it energy} is one of the most fundamental
concepts of
physics and its conservation is evident in all its subdisciplines.
This is why we apply the principle of energy conservation to the
motion of the test particle
in gravitational field
by requiring that its energy,
gained by gravitation, has to be balanced by the decrease of
its mass.
Consider, for
simplicity,
the static spherically symmetrical gravitational
field of a large mass $M$ in which a test particle $m$ occurs at
distance $r$ from $M$.
We demand that the sum of the inertial {\it
as well as}
the gravitational mass of the test particle at a spatial
point $r$ and the work being done by its adiabatic translation
($v\approx 0$) from
infinity to $r$ must be equal to its rest mass $m_{\infty}$
in infinity, i.e.
$$m_{\infty}=m(r)+1/c^2\int_{\infty}^r F(x)dx,\quad\eqno(a)$$
where $F(r)$ is the force acting on the test particle in gravitational field.
Inserting Newton's law into Eq.(a) one obtains
$$m(r)=m_{\infty}\gamma(r) =m_{\infty}\exp{(-\lambda/r)};\quad
\lambda=\frac{GM}{c^2}\quad\eqno(b)$$
and the new modified Newton's force law reads
$$F_N(r)=-\frac{GMm_{\infty}\exp{(-\lambda/r)}}{r^2}.\quad\eqno(c)$$
The factor
$\exp{(-\lambda/r)}$, appearing in the Newton-like force law, has a
remarkable consequence, namely when the test particle is approaching
sufficiently near
to
a (point-like) gravitating mass, the force
acting on it becomes weaker and weaker till it completely ceases at
the gravity center. On the other side, $F_N(r)$ coincides asymptotically
with common Newton's law in the regions where the field is weak
($\lambda\ll 1$).

In our approach, the particle mass in gravitational field apparently
depends on its position. On the other side,
it is generally assumed that particle rest mass is, in classical
gravity as well as in general
relativity, constant.
This assumption, as stressed by Dicke \cite{DIC}, is very partially
supported by
experiment. Especially, this constancy has never been tested in strong
gravitational fields.
The concept of variable rest mass, motivated by different reasons, is not new.
It appears, e.g. in Dicke reformulation \cite{D} of
Brans-Dicke theory \cite{BD}, in Hoyle and Narlikar's theory of
gravitation \cite{HO}, in Malin \cite{MA} cosmological theory of variable
rest masses,
in Beckenstein theory of rest mass field \cite{BE}
and in Vondr\'a\v cek \cite{V} Mach's theory of gravitation,
to mention only
a few.

Since at each spatial point of gravitational field the force acting
on the test particle is given by Eq.(c) one can consider the
neighborhood of $M$ as a classical force field, i.e. one can
determine its intensity, its potential, its field energy density,
and its total field energy. For the Newton-like field, all these
field quantities assume everywhere finite and positive values. As a
consequence, the energy conservation, the total energy released by
any gravitational collapse can not be larger than $nm_{\infty}c^2$,
$n$ being the number of particles participating on the gravitational
collapse. Since the energy released by a gravitational collapse is
restricted no naked singularity, in classical sense, can be formed
in nature. Moreover, a star-like objects that is larger than
approximately $2M_{sun}$-Chandrasekhar limit- might exist in
equilibrium due to weakening of gravitation toward the gravity
center (see \cite{GM}).

There is a link between the Newton-like gravitation theory and the
geometrical structure of spacetime. To show this we consider a
spherically symmetrical, asymptotically flat gravitation field with
local Lorentz reference frame co-moving with the freely falling
elevator. The velocity of this system is changing  and, accordingly,
the corresponding relativistic length contraction and time
dilatation. Therefore, one can ascribe to each spatial point of
gravitational field the special-relativistic factor
$\gamma=\sqrt{1-v^2/c^2}$. Knowing this factor as a function of
position $r$ the line element in the reference frame of rest
observer turns out to be
$$ds^2=\gamma^{-2} dr^2+r^2(d\theta^2 +\sin^2 \theta
d\phi^2)-\gamma^{2}c^2dt^2,\qquad \gamma=\sqrt{1-\frac{v^2}{c^2}}
\quad\eqno(d)$$
The special-relativistic energy conservation for a
freely falling particle leads to the formula
$$\frac{m_{\infty}c^2\gamma(r)}{\sqrt{1-v^2/c^2}}=m_{\infty}c^2,\quad\eqno(e)$$
where $m_{\infty}$ is the mass of particle in infinity. This formula,
which tell us that the total energy at each spatial point of
gravitational field is constant
equal to $m_{\infty}c^2$,
represents a bridge between mass decrease function (MDF, for short)
$\gamma(r)$ of a particle in
gravitational field and its geometrical structure.
Inserting MDF of the Newton-like field
$\gamma(r)=\exp{(-\lambda/r)}$ into Eq.(e)
then the line element (d)
becomes the form
$$ds^2=\exp{(\frac{2\lambda}{r})} dr^2+r^2(d\theta^2 +\sin^2 \theta
d\phi^2)-\exp{(-\frac{2\lambda}{r})}c^2dt^2.\quad\eqno(f)$$
Apparently, this line element coincides for weak field ($\lambda/r\ll 1)$
asymptotically
with the Schwarzschild one.

The metric, associated with the Newton-like
field, represents a non-vacuum solution of Einstein's
equations.
In order to determine the source terms assigned to this metric we
 solved the inverse problem of general relativity, i.e. to
find the source terms given the metric.
The (tt)-component of
energy-momentum, assigned to the metric (f) represents
a regular function of $r$ assuming at $r=0$ a finite value.

As is common knowledge, gravitationally collapsing object of sufficient
mass are doomed to form black holes defined by an event horizon within
which resides the singularity of general relativistic equations.
Characteristic feature of the metric associated with the Newton-like
field is that, in the classical black hole model, the surface of
events and the infinite red shift are shifted into center of a
collapsing mass, which implies the non-existence of common
black-hole. Hence, strict
application of energy conservation has at least to important
consequences: the impossibility of forming the naked singularity
and the non-existence of black holes.

It appears that the 'field' and 'geometrical' aspects of
gravitation are mutual related. Given MDF one can determine the
geometry of spacetime where the gravitational field occurs, and
given the geometry one can determine the corresponding MDF and so its field
quantities.

The organization of this article is as follows.
In Section 2, we derive MDF and the modified force
law from the requirement of the constancy of the
energy content of the test particle during its motion in gravitational
field.
In Section 3, we describe properties of the Newton-like field which
arises if one inserts into Eq.(a) the common Newton force law.
In Section 4, we determine the metric associated with the Newton-like
field and calculate the source terms of Einstein's equations by inserting
components of metric tensor into left-hand sides of these equations.
In Section 5, we study the actual gravitational fields and metrics
associated with them. In section 6, we  put forward the hypothesis
concerning
gravitation as a force field embedded into metric
determined by
the given mass decrease function.

\section{The Mass as a Function of Gravitational Field }

In electrostatics, one derives an expression for the energy density of
the electrostatic field by calculating the work done in assembling a charge
distribution from elements of charge that are initially in a dispersed
state. A similar situation in classical Newton theory leads to a strange
conclusion that the energy density of gravitational field is {\it
negative}
definite (see, e.g. \cite{U}).
In the Newton theory, the gravitational energy stored in a system
of masses can be found by calculating the work done in bringing these
masses from infinity to the final position. For a system of two
masses $m_1$ and $m_2$ we get the gravitational potential energy by the
familiar formula
$U(r)=-Gm_1m_2/r,$
where $r$ is the distance between the masses. An interesting feature of
classical gravity is the possibility that in  certain situations the {\it
total}
energy of a gravitating system can become {\it negative} supposing the
constancy of its masses. As an example,
consider a sphere of gas in convective equilibrium which radiates away
its excess thermal energy as it slowly contracts. Its total energy is
\cite{I}
$$E=Mc^2-\frac{6}{7}\frac{GM^2}{r}.$$
We see that up a certain $r$ the total energy of this system becomes
negative.
In the Maxwell-like field theory of gravitation,
it is assumed that the total energy content of gravitating mass system
consists of positive
energy of its masses and {\it negative} field energy so that the total
energy
appears to be constant (for details see, e.g. \cite{U}).
However, since the energy is by definition positive
the gravitating system with the {\it total}
negative energy is in principle inadmissable.

The simplest way how to avoid this situation and to
conserve the total energy of a gravitating
mass system, without introducing the negative field energy, is to
assume that the work done by a moving
test particle  in the gravitational field
is going at the expense of
its internal energy, i.e.
\begin{equation} \label{1}
m(r)=m_{\infty}-
\frac{1}{c^2}\int_{\infty}^r F(x)dx= m_{\infty}-
\frac{1}{c^2}\int_{\infty}^r m(x)\chi(x)dr,
\end{equation}
where $m_{\infty}$ is the mass of particle in infinity and $\chi(r)$
is the the so-called force function.
Writing $m(r)$ and $F(r)$ in the form $m_{\infty}\gamma(r)$ and
$F(r)=m(r)\chi(r)$,
and differentiating
Eq.(\ref{1}) with respect to $r$, we get
the following differential equation (a prime means differentiation with
respect
to $r$)
$$\gamma(r)'=-\frac{\gamma(r)\chi(r)}{c^2}$$
whose solution
\begin{equation} \label{3}
\gamma(r)=\exp{\left(-\frac{1}{c^2}\int_{\infty}^r{\chi(x)dx}\right)}
\end{equation}
gives the relation between $\chi(r)$ and $\gamma(r)$.
We remark that the inertial and the gravitational mass are in
gravitational field
equally changed so that its {\it ratio} remains constant and
the principle of
equivalence
is not violated.

Given $\gamma(r)$ or $\chi(r)$ one can derive several important field and
characteristics of gravitational field.
Using the familiar formula, $F(r)=\frac{dE(r)}{dr}$, and taking into
account that the energy content of a particle is $m(r)c^2$,
the force acting on this particle in  gravitational field is
\begin{equation} \label{4}
F_{g}(r)=-\frac{d(m(r)c^2)}{dr}=- m_{\infty}\chi(r)
\exp{\left(-\int_{\infty}^{r}{\frac{\chi(x)dx}{c^2}}\right)}
=-m_{0}\gamma(r)'c^2.
\end{equation}
The potential energy of $m_{\infty}$ reads as ($\gamma(\infty)=1$)
$$E_{pot}=-\int_{\infty}^{r}F(x) dx=m_{\infty}c^2(\gamma(r)-1).$$
The spatial point $r_0$ at which the rest energy of $m_{\infty}$
is totally exhausted is
given by the equation
\begin{equation} \label{5}
\frac{1}{c^2}\int_{\infty}^{r_0}F_g(x) dx=m_{\infty}.
\end{equation}

In the next section we will apply the above formulas to a type
of gravitational field called the Newton-like force field.

\section{Field Properties of the Newton-like Gravitational Field}

The field that arises if one inserts
into Eq.(\ref{1}) the Newton force law we obtain
{\it the Newton-like} field.
By means of
Eq.(\ref{1})),
one gets for MDF of this field
the expression $\gamma_N(r)=\exp{(-\lambda/r)}$, so that it holds
$$m_{N}(r)=m_{\infty}\gamma_N(r)=m_{\infty}\exp{\left(-\frac{\lambda}{r}\right)};\qquad
\lambda=GM/c^2.$$
The Newton-like force law follows from Eq.(3)
\begin{equation}
F_{N}(r)=
-\frac{GMm_{\infty}}{r^2}
\exp{\left(-\frac{\lambda}{r}\right)}=
-\frac{c^2\lambda m_{\infty}\exp{(-\lambda/r)}}{r^2}.
\end{equation}
The corresponding potential $U_N(r)$ reads as
\begin{equation}\label{pot}
U(r)=-\int_{\infty}^{r}{F(r)}dr=m_{\infty}c^2(\exp{(-\lambda/r)}-1).
\end{equation}
The graph of $F_N(r)$ as a function of $r$ is depicted in Fig. 1.\\
\begin{figure}[h]
\includegraphics[scale=0.8]{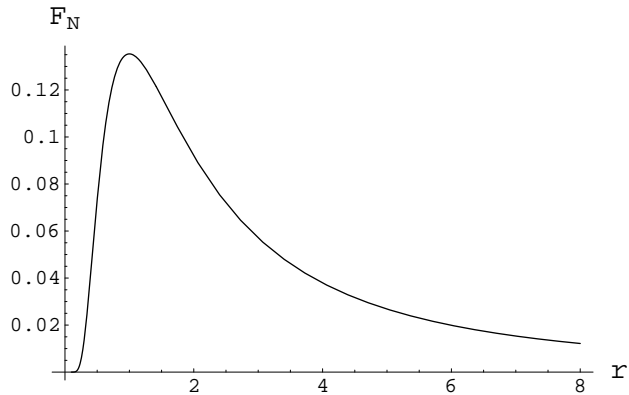}
\caption{The Newton-like force $F_N$ as a function
of $r$ ($\lambda=G=c=m_{\infty}=1$}. \label{fig1}
\end{figure}

$F_{N}(r)$ represents  everywhere finite
function assuming at $r=0$ and $r=\infty$ zero value.
The maximal value of $F_N(r)$ is reached at $r_{max}=GM/2c^2$ where it
assumes
the finite value
$$F_{N}^{(max))}=\left
(\frac{c^4}{G}\right)\left(\frac{m_{\infty}}{M}\right)(4\exp{(-2)}).$$
$F_N^{(max)}$ depends on the quantity $c^4/G$, the ratio
$m_{\infty}/M$ and a numerical factor $4\exp{(-2)}\approx 1$. The
quantity $c^4/G$ is related to the other Planck relativistic
constants (Planck mass $m_P$, Planck-Wheeler's length $l_P$ and
Planck time $t_p$) through the equation
$$\frac{c^4}{G}=\frac{hc}{l_{p}^2}=m_pl_p(t_p)^{-2}.$$
Note that $c^4/G$, though being one of Planck's relativistic constants
arisen by the combination
of $G$, $c$ and $h$, {\it does} not contain $h$.\\
Eq.(6) we rewrite in a symmetrical form
\begin{equation} \label{5a}
F_{(N)}(r)=-\frac{\sqrt{G}m_{\infty} \sqrt{G}M}{r^2}\exp{(-\lambda/r)},
\end{equation}
where $\sqrt{G}m_{\infty}$ and $\sqrt{G}M$ are the so-called gravitational
charges
of $m_{\infty}$ and $M$ \cite{IS}.
The force acting on one unit of gravitational charge $\sqrt{G}m_{\infty}$
becomes
field intensity of the Newton-like gravitational field
\begin{equation}\label{6}
I_{(N)}(r)=-\frac{\sqrt{G}M}{r^2}\exp{\left(-\frac{\lambda}{r}\right)}.
\end{equation}
For the field energy/mass density of the Newton-like field we get the
expression
\begin{equation}\label{7}
E_N(r)=\frac{I_N(r)^2}{8\pi}=\frac{GM^2}{8\pi r^4}\exp{\left(-\frac{2\lambda}{r}\right)}\quad {\rm
and}\quad \rho_{N}(r)=\frac{GM^2}{8\pi
r^4c^2}\exp{\left(-\frac{2\lambda}{r}\right)},
\end{equation}
respectively.
The graph of $E_N(r)$ as a function of $r$ is depicted in Fig. 2.\\

\begin{figure}[h]
\includegraphics[scale=0.8]{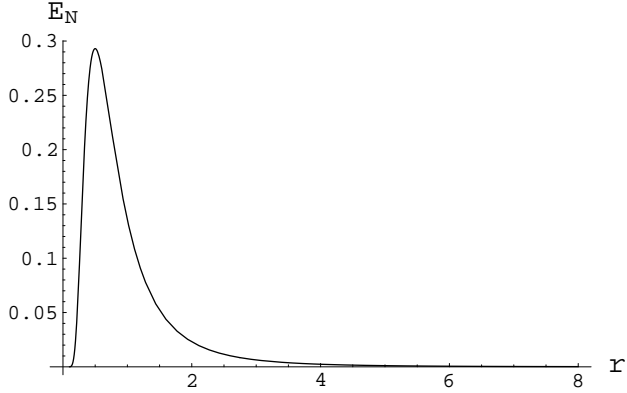}
\caption{The energy density of the Newton-like field $E_N$ as a
function of $r$ ($\lambda=G=M=1$)}. \label{fig2}
\end{figure}

Again, $E_N(r)$ represents an everywhere {\it positive} and finite function.
It is interesting that
the total energy of the Newton-like
gravitational field of a point-like $M$ is
a finite value.
$${\bf E}_{N}=\frac{1}{2}\int_{\infty}^0{ \frac{GM^2}{r^4}
\exp{\left(-\frac{2\lambda}{r}\right)} r^2} dr=\frac{Mc^2}{4}.$$
The work done by adiabatic
translation a test particle from infinity to the gravity center in
the Newton-like field is
just equal to its whole internal energy content
$$\int_{\infty}^{0}F_{N}(r) dr=m_{\infty}c^2.$$

Next we show that MDF and
the space-time geometry of a static static central symmetric gravitational
field are related.

\section{Metric in a Freely Falling Elevator}

Consider a spherically symmetrical and asymptotically flat
gravitational field with the local Lorentz reference frame comoving with
the freely falling elevator
from infinity  toward the central body $M$.
In this elevator, one feels no
gravitation field, because it carries the Euclidean metric
supposed to be in infinity.
In distance $r$ from the central body
the elevator moves with velocity $v$.
Relativistic length contraction and time dilatation in the reference
frame of the central body is changing in accord with the velocity of the
elevator.
The line element in the reference frame of elevator ($\Sigma_1$) is
$$ds^2=dr^2+r^2(d\theta^2 +\sin^2(\theta) d \phi^2)
-c^2dt^2,$$
while in that of the central body ($\Sigma_2$),
due to the Lorentz transformation, becomes
$$ds^2=\gamma^{2} dr^2+r^2(d\theta^2 +\sin^2 \theta
d\phi^2)-\gamma^{2}c^2dt^2,\qquad \gamma=\sqrt{1-\frac{v^2}{c^2}}.$$

The principle of the energy conservation applied to a
freely falling particle in gravitational field, supposing that
$E_{tot}=0$ in infinity,
leads to the equation
\begin{equation} \label{energy}
E_{tot}=E_{kin}+E_{pot}=\frac{m_{\infty}c^2\gamma(r)}{\sqrt{1-v^2/c^2}}-m_{\infty}\gamma(r)+
m_{\infty}c^2(\gamma(r)-1)=0
\end{equation}
From Eq.(\ref{energy}) follows
\begin{equation} \label{energy1}
\frac{m_{\infty}c^2\gamma(r)}{\sqrt{1-v^2/c^2}}=m_{\infty}c^2,\quad
\gamma(r)=\sqrt{1-\frac{v^2}{c^2}},
\end{equation}
i.e. the energy content of a test particle during its moving in
gravitational field remains constant equal to $m_{\infty}c^2$.

This is an important equation that expresses energy conservation of a
particle in gravitational field. It represents also a bride
between the spacetime geometry
and field properties of a gravitational field.
Given $\gamma(r)$ we can determine the metric of
every static spherically symmetrical field.
Likewise, given the metric
one can determine the
corresponding MDF for every static spherically symmetrical field.
 $\gamma_{N}(r)$ of the Newton-like field implies
the metric
$$g_{rr}=\gamma(r)^2=
\exp{(2\lambda/r)}\qquad g_{tt}=-(\gamma)^{-2}=-
\exp{(-2\lambda/r)}.$$
The line element associated with the Newtonian MDF is
\begin{equation} \label{12}
ds^2=-\exp{(-\frac{2\lambda}{r})}dt^2+\exp{(\frac{2\lambda}{r})}dr^2 + r^2(d\theta^2 +\sin^2 \theta
d\psi^2).
\end{equation}
This line element  asymptotically
coincides with the
Schwarzschild one for weak field $\lambda\ll 1$. The metric (\ref{12})
describes a
non-vacuum solution of Einstein's equations with non-zero
source terms.

As is well-known, the line element (metric) of a static, spherically
symmetric spacetime
can be written in the form
\begin{equation} \label{R}
ds^2=-\exp{(2B(r))} dt^2 +\exp{(2A(r))}dr^2+r^2(d\theta^2+\sin^2{\theta}
d\psi^2),
\end{equation}
where
the standard spherical coordinates of a distant observer are used (we
implicitly suppose the asymptotic flatness of the spacetime).
The non-null components of the
standard Einstein tensor $G_{\nu\mu}$ read as (G=c=1)
\begin{equation} \label{aa}
G_{tt}=\frac{\exp{(2B)}}{r^2}\frac{d}{dr}\left
(r[1-\exp{(-2A)}]\right)
\end{equation}
\begin{equation}  \label{bb}
G_{rr}=-\frac{\exp{(2A)}}{r^2}[1-\exp{(-2A)}]+\frac{2}{r}\frac{dB}{dr}
\end{equation}
\begin{equation} \label{cc}
G_{\theta\theta}=r^2\exp{(-2A)}\left [\frac{d^2B}{dr^2}+
(\frac{dB}{dr})^2+
\frac{1}{r}\frac{dB}{dr}-\frac{1}{r}\frac{dA}{dr}-
\frac{dA}{dr}\frac{dB}{dr}\right ]
\end{equation}
\begin{equation}\label{dd}
G_{\phi\phi}=\sin^2{\theta}G_{\theta\theta}
\end{equation}
Comparing Eq.({\ref{R}) and Eq.(\ref{12}) we find that
$A(r)=\lambda/r$ and $B(r)=-\lambda/r$. With $A(r)$ and $B(r)$, Eqs.(\ref{aa}), (\ref{bb}), (\ref{cc})
and (\ref{dd}) turn out to be
$$G_{tt}=\frac{\exp{(-2\lambda/r})}{r^2}Q(r);\quad
G_{rr}=-\frac{\exp{(2\lambda/r)}}{r^2}Q(r),$$
where
$$Q(r)=1-\exp{(-2\lambda/r)}(1+\frac{2\lambda}{r})$$
and
$$G_{\theta\theta}=\frac{2\lambda^2}{r^2}\exp{(-2\lambda/r)} \qquad
G_{\phi\phi}=\frac{2\lambda^2}{r^2}\exp{(-2\lambda/r)}\sin^2\theta.$$
$Q(r)$ represents a monotonous decreasing function of $r$ that for
$r\rightarrow 0$ and $r\rightarrow \infty$ assumes the limit value $Q(r)\rightarrow 1$ and
$Q(r)\rightarrow 0$, respectively.
$Q(r)$ as a function of $r$ is depicted in Fig.3.\\

\begin{figure}[h]
\includegraphics[scale=0.8]{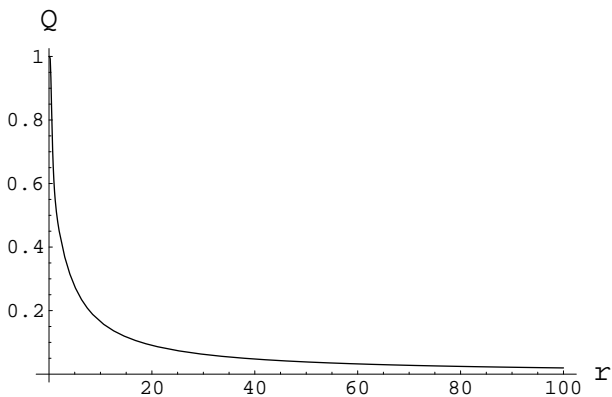}
\caption{The quantity $Q(r)$ as a function of $r$.} \label{fig3}
\end{figure}

For weak field $\lambda/r\ll 1$, one can set $\exp{(-2\lambda/r)}\approx
1-2\lambda/r$ which inserting in $Q(r)$ yields
$Q(r)=-4\lambda^2/r^2$ and we have
\begin{equation}
G_{tt}\approx -\frac{4\lambda^2\exp{(-2\lambda/r)}}{r^4}=8\pi G
\left(\frac{4GM^2\exp{(-2\lambda/r)}}{8\pi r^4}\right)
=8\pi G (4 \rho_M(r)).
\end{equation}
The results concerning the non-vacuum spacetime associated with
the Newton-like field
can be summed up as follows:\\
(i) $G_{tt}(r)$ represents everywhere regular function assuming at
the center of gravity zero value. The graphical representation of
$G_{tt}(r)$ as a function of $r$ $(\lambda=1)$ is shown in Fig.4.\\
\begin{figure}[h]
\includegraphics[scale=0.8]{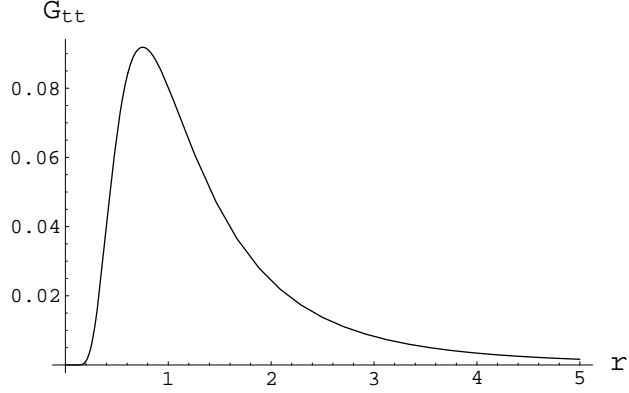}
\caption{The (tt)-component of Einstein tensor as a function of $r$
($\lambda=1$). } \label{fig4}
\end{figure}
The graph despises a one-hump curve which reaches its maximal value
at $r\approx 0.7$ and then asymptotically
decreases to zero.\\
(ii) The integrals of $G_{tt}(r)$,$G_{\theta\theta}(r)$ and
$G_{\phi\phi}(r)$
over the interval $r\in [0,\infty]$ assume finite values
($G_{tt}(r)\approx 1.2\lambda,$
$G_{\theta\theta}(r)=G_{\phi\phi}(r)=1\lambda$.\\
(iii) Comparing graphs of
$G_{tt}(r)$ and $\rho_N(r)$ (Fig.5) we see
that they have practically the same shapes which point out
that $G_{tt}$ is essentially given by the energy density of
the Newton-like field.  \\
\begin{figure}[h]
\includegraphics[scale=0.8]{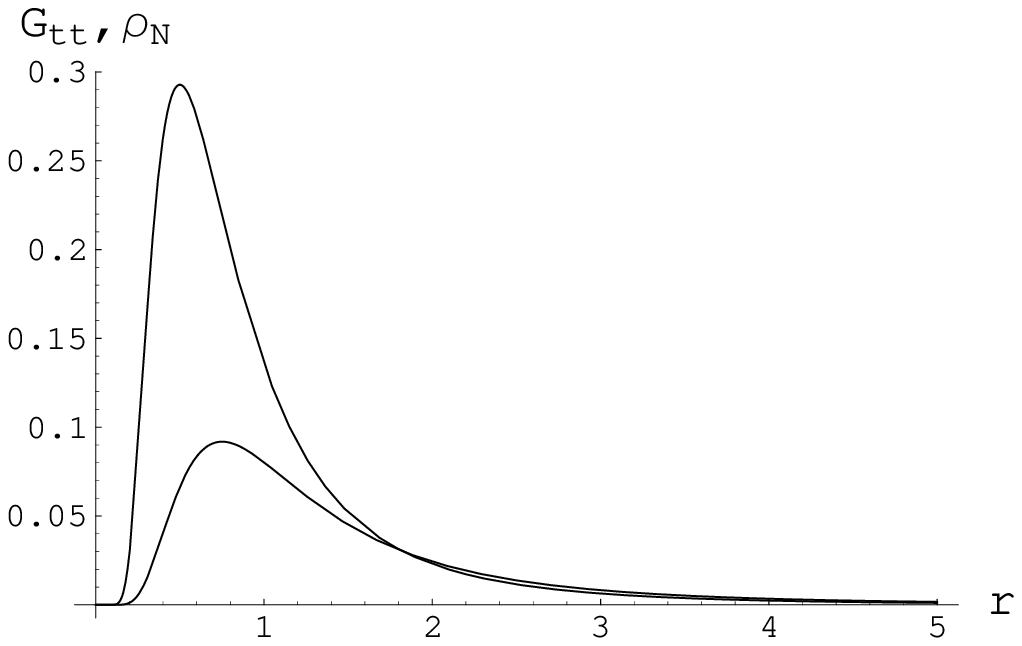}
\caption{$G_{tt}$ (down) and $\rho$ (up) as a function of $r$
($\lambda=1$)}.\label{fig5}
\end{figure}
(iv) For the weak field $\lambda/r\ll 1$, $G_{tt}(r)$ is, up to a constant,
equal to
$\rho_N(r)$.\\

The components of the Ricci tensor for the metric
$g_{rr}=\exp{(2\lambda/r)}$ and $g_{tt}=\exp{(-2\lambda/r)}$
assume the forms
$$R_{rr}=\frac{2\lambda^2}{r^4}\quad
R_{tt}=-\frac{2\lambda^2\exp{(-4\lambda/r)}}{r^4}\quad
R_{\theta\theta}=-1+\frac{\exp{(-2\lambda/r)}(2\lambda+r)}{r}$$
and
$$R_{\theta\theta}=R_{\theta\theta}\sin^2\theta.$$
It is straightforward to verify that
$$R_{tt}g_{rr}=-R_{rr}g_{tt}=8\pi G\left (\frac{GM^2}{8\pi
r^4}\exp{(-\frac{2\lambda}{r})}\right)=8\pi G\rho_N(r),$$
where $\rho_M(r)$ is exactly the field energy density of the Newton-like field.

In an almost empty flat space,
where the gravitational field is relatively weak, the source of
gravitational field is its own field energy density
$E_{g}(r)=I^2_g(r)/8\pi$.
Inserting $E_g(r)$ into the familiar differential equation we get
$$
\frac{dI_{g}(r)}{dr}+\frac{2}{r}I_g(r)=-\frac{1}{2\kappa}E_g(r)=-\frac{1}{2\kappa}
I_g^2(r),$$
where $\kappa$ is a phenomenological constant.
 \footnote{Strictly speaking the energy
density of gravitational field is, in
analogy to electrostatics, given as
$E_g(r)=D_g(r)I_g(r)/8\pi,$ where
$I_g(r)$ is the intensity and $D_g(r)$ the displacement of gravitational
field. The intensity and displacement
are related by the equation  $D_g(r)=\epsilon_g
E_g(r)$ where $\epsilon_g$ is gravitational
'dielectric' constant. Therefore, the energy density of gravitational
field can be written as $E_{g}(r)=\epsilon_{g}I_g(r)^2$ and $1/\kappa$ may
be interpreted as the gravitational 'dielectric' constant for the considered
medium.}
This differential equation has for a point-like mass $M$ a simple
general solution
$$I_g(r)=-\frac{C}{r^2}-\frac{2\kappa}{r}.$$
Setting $C=GM$ we get
\begin{equation}  \label{gal}
I_g(r)=T_1(r)+T_2(r);\quad T_1=-\frac{GM}{r^2}\quad T_2=-\frac{2\kappa}{r}.
\end{equation}
In the region where the first term $T_1(r)$ prevails ($r\ll 2\kappa$) the
intensity of
gravitational field follows practically Newton's law while in the
region where the second term $T_2(r)$ prevails ($r\gg 2\kappa$) we get a modified
force law proportional to $1/r$.
Taking the simplest model of a spiral galaxy as a point-like mass
concentrating
in the galactic core then, according
to Eq.(\ref{gal}), the most part of gravitational energy
is concentrated in its halo.
Rotation curves
in the vicinity of the galactic core are Keplerian
while those in the region where the second term
$T_2\propto 1/r$ is prevailing
become asymptotically flat. It is noteworthy that when setting
$\kappa=\sqrt{GMa_0}$, $a_0$ being the Milgrom acceleration constant,
we get similar formula for gravitation force
as in the successful formalism of MOND which describes
regularity in the formulation and evolution of galaxies (see, e.g.
\cite{SAN}).
Therefore, it is tempting to interpreted the gravitational energy as one of
the
components of dark matter.

\section{Force fields and their metrics}

As is well known, the Reissner-Nordstr\"om metric represents a
non-vacuum solution of Einstein's equations. The spacetime around
the mass with charge $Q$ (in e.s.u.), is, in static spherically
symmetrical case, described by the metric
$g_{tt}(r)=-(1-2GM(r)/r+Q^{2}/c^2r^2)$ and
$g_{rr}(r)=(1-2GM(r)/r+Q^{2}/c^2r^2)^{-1}$. $M(r)$ is the mass
interior to radius $r$. The third term of the metric stems from the
energy density of the electrostatic field. If we take, instead of
the energy density of the electrostatic field ($E(r)=Q^2/r^4$), the
energy density of the Newtonian field as the first approximation to the
energy density of gravitational field, i.e. $E(r)=GM^2(r)/r^4$, then
we obtain a modified Reissner-Nordstr\"om metric for a neutral
gravitating mass $M$: $g'_{tt}(r)=-(1-2GM(r)/r+GM(r)^2/c^2r^2)$ and
$g'_{rr}(r)=(1-2GM(r)/r+GM^2/c^2r^2)^{-1}$. Here, the third term in
$g'_{tt}(r)$ and $g'_{rr}(r)$ stems from the energy density of the
Newtonian force field. If the gravitating mass $M$ is isolated then
the modified Reissner-Nordstr\"om metric becomes a simple form
$g_{tt}(r)=-(1-\lambda/r)^2$ and $g_{rr}(r)=(1-\lambda/r)^{-2}$. MDF
assigned to the modified Reissner-Nordstr\"om field is
$\gamma_{RN}(r)=-(1-\lambda/r)$ which implies Newton's law. The
intensity and field energy density of Newton's field are shown in
Table 1 (the second colon). Neglecting $\lambda^2/r^2$ in
$g'_{tt}(r)$ and $g'_{rr}(r)$, i.e. neglecting any energy density of
gravitational field, the Reissner-Nordstr\"om field reduces to
Schwarzschild one whose MDF is $\gamma_{S}(r)=\sqrt{1-2\lambda/r}$
\footnote{$\gamma_S(r)$ is identical with the 'Bardeen' potential
introduced by the investigation of black holes
 \cite{BA}.}
and the corresponding force law have the form
$$F_S(r)=\frac{GMm_{\infty}}{r^2\sqrt{1-2\lambda/r}}.$$
Table 1 shows properties of various fields derived from the
corresponding MDFs. The intensity and energy density for
Schwarzschild's field are  shown in Table 1 (third colon). Last
field we considered is that whose MDR is identically equal to $1$
(or generally to a constant). This means that the mass in this field
{\it does not} change at all and, as a consequence of it, its force
is identical equal to zero (see Table 1 fourth colon). In Table 2,
we present the metrics assigned to the above force fields.
\begin{center}
\begin{tabular}{|c|c|c|c|c|}
\hline
field & MDF &field intensity & energy density &
total energy\\
\hline
Newton-like &-$\exp{(-\lambda/r}$ &$ -\lambda\exp{(-\lambda/r)}/r^2$&
$\lambda^2\exp{(-2\lambda/r)}/8\pi r^4$ &$ M/4 $\\
Mod. R-N&$-(1-\lambda/r)$ &$-\lambda/r^2$ & $\lambda^2/8\pi r^4$ &$ \infty$\\
Schwarzschild&$-\sqrt{1-2\lambda/r}$ &$-\lambda/r^2\sqrt{1-2\lambda/r}$ &
$\lambda^2/(8\pi r^4(1-2\lambda/r))$  & $\infty$\\
flat &1 & 0 & 0 & 0\\
\hline
\end{tabular}\\
\end{center}
\begin{center}
Table 1
\end{center}

\begin{center}
\begin{tabular}{|c|c|c|c|c|}
\hline
 &$g_{tt}$ &$g_{rr}$ & spatial point $r_0$&
$G_{tt}$\\
\hline
Newton-like &$-\exp{(-2\lambda/r)}$&$\exp{(2\lambda/r)}$ &$0$ &$
(\exp{(-2\lambda)}Q_{tt})/r^2 $\\
Mod. R-N &$-(1-\lambda/r)^2$ & $(1-\lambda/r)^{-2}$ &$ \lambda$ &$
(\lambda^2(1-\lambda/r)^2/r^4$\\
Schwarzschild &$-(1- 2\lambda/r)$ &
$(1-2\lambda/r)^{-1}$  &$2\lambda$ & $0$\\
flat & 1 & 1 & - & 0\\
\hline
\end{tabular}\\
\end{center}
\begin{center}
Table 2
\end{center}
\section{Concluding Remarks}
Taking into account what has been said so far we put forward the
following hypothesis concerning gravitation. We assume that the
neighborhood of a gravitating object is the space where the rest
mass of a particle is transformed into its kinetic energy, according
to Einstein's equivalence principle between mass and energy, whereby
its total energy remains conserved. Due to transformation of mass
into energy the velocity of particles in gravitational field are
changed and with this change also the local Lorentz factor and the
metric. Spacetimes in regions, where gravitation occurs, appears to
rest observer generally as curved. Their metrics are determined by
the density of the ordinary matter as well as by the energy density
of the force fields. Gravitational field appears as a force field
embedded into metric determined by its mass decrease function.

The presented picture of gravitation has at least two important consequences:\\
(i) {\it In classical gravitation theory}.
When a star run out of
nuclear fuel, the only force left to sustain it against gravity is the
pressure associated with the zero-point oscillation of its constituent
fermions. This is valid if the gravitational force obeys Newton's
law. If one take, instead of Newton's, the Newton-like
force law then the force is weakening
near the gravity center as a consequence of the exhausting
of particle rest mass.
Due to principle
of the energy conservation, therefore the
total energy released by the gravitational collapse
can not be larger than the sum of all masses participating at it.
The strict application of energy conservation
prevent so the forming of the naked singularity.
The energy balance for an isolated collapse (nothing is put into or
taken out of it), i.e. energy release plus the internal energy of
masses has to be during the whole process constant. On the other
side, the weakening of force  near to the gravity center in the
Newton-like field  makes it possible the existence of star-like objects in
equilibrium which are more collapsed than neutron stars (see \cite{GM}.
The work
done by translation of a test particle from infinity to the center
of gravity is, in contrast to the Schwarzschild field, for the
Newton-like field just equal to its rest mass in infinity.\\
(ii) {\it In the general relativity}.
The strict application of energy conservation yields an important relation
between a gravitational force field and the associated spacetime geometry.
The metric associated with the Newton-like field is remarkable simple
$g_{tt}=\exp{(-2\lambda/r)}$ and $g_{rr}=\exp{(2\lambda/r)}$.
It coincides for weak field ($\lambda/r\ll 1)$ asymptotically
with the Schwarzschild metric and represents a non-vacuum solution of
Einstein's equations.
The (tt)-component of
energy-momentum represents assigned to this metric represents
a regular function of $r$ assuming at $r=0$
a finite value. Its integral over the whole range of radius $r$ is
 likewise
a finite number. In the common black hole model, the metric
associated with the Newton-like field shifts the surface of an
infinite red shift and the horizon of events into the center of
gravity, which implies the non-existence of the 'classical' black
holes.\\

It is noteworthy that the dependence of mass on position
satisfies one requirement of Mach' principle, namely
that that mass of particle should depend on neighboring mass
distribution (see, e.g. \cite{N},\cite{E}).

The application of the principle of energy conservation to the motion
of particle in gravitational field
leads to a modified force law and implies a remarkable simple metric
with the important
consequences. It prevents the naked singularity and leads to
a metric preventing also the forming of classical black holes.
The energy of gravitational field can be interpreted as one component of
dark matter.\\

The aim of this article was only to outline the basic ideas
concerning the strict application of energy conservation in the
gravitation theory, therefore
everything is simplified and
many important issues remained open which will be subject of
subsequent works.

\end{document}